# Gravitational oscillations in multidimensional anisotropic model with cosmological constant and their contributions into the energy of vacuum

Sergey Yakovlev
sergey.y@mail.ru

**Moscow, November 24, 2011**

Were studied classical oscillations of background metric in the multidimensional anisotropic model of Kazner in the de-Sitter stage. Obtained dependence of fluctuations on dimension of space-time with infinite expansion. Stability of the model could be achieved when number of space-like dimensions equals or more then four. Were calculated contributions to the density of «vacuum energy», that are providing by proper oscillations of background metric and compared with contribution of cosmological arising of particles due to expansion. As it turned out, contribution of gravitational oscillation of metric into density of «vacuum energy» should play significant role in the de-Sitter stage.

Anisotropic metric for the case of three dimensional space-hypersurface was investigated by Edvard Kazner in 1992, and appeared possibility to investigate effects of anisotropy in Einstein's equations.

Like typical task is interesting the investigation of stability of such model, and describing the little fluctuations of the metric. Gravitational waves on the background of curved space-time are describing by the equation

$$\bar{h}_{\mu\nu|\alpha}{}^{\alpha} + 2R_{\alpha\mu\beta\nu}\bar{h}^{\alpha\beta} = 0, \qquad (1)$$

where covariant derivative with vertical line is taking in the unperturbed background metric, which we'll denote $g_{\mu\nu}(x)$, and correspondingly full metric is given by

$$g_{\mu\nu}^{full}(x) = g_{\mu\nu}(x) + h_{\mu\nu}(x),$$

rising up and dropping down of indexes is doing by background metric tensor. Here

$$\bar{h}_{\mu\nu} = h_{\mu\nu} - \frac{1}{2}hg_{\mu\nu},$$
$$h = h_{\alpha}^{\alpha}.$$

In Eq. (1) is supposed, that we've imposed Lorentz gauge on the metric

$$h_{\mu\alpha}{}^{|\alpha} = 0. \qquad (2)$$

Let's consider the model of multidimensional space-time without substance with cosmological term. Einstein's equations are writing down in the form

$$R_{\alpha\beta} - \frac{1}{2}g_{\alpha\beta}R = -\Lambda g_{\alpha\beta}, \qquad (3)$$

or

$$R_{\alpha\beta} = \frac{2\Lambda}{n-1}g_{\alpha\beta},$$

where $n$ is number of space dimensions.
Eq. (1) in the first order of perturbation [6] is transforming into

$$\bar{h}_{\mu\nu|\alpha}{}^{\alpha} + 2R_{\alpha\mu\beta\nu}\bar{h}^{\alpha\beta} - 2R_{\alpha(\mu}\bar{h}_{\nu)}{}^{\alpha} = \frac{2\Lambda}{n-1}h_{\mu\nu}, \qquad (4)$$

where little round brackets with indexes denote symmetrization. Calculating the third term in the right side of Eq. (4), we get

$$2R_{\alpha(\mu}\bar{h}_{\nu)}{}^{\alpha} = \frac{2\Lambda}{n-1}(g_{\alpha\mu}\bar{h}_{\nu}{}^{\alpha} + g_{\alpha\nu}\bar{h}_{\mu}{}^{\alpha}) = \frac{4\Lambda}{n-1}\bar{h}_{\mu\nu}.$$

We'll use following expression for the square of space-time interval

$$ds^2 = -dt^2 + e^{2\alpha(t)+2\beta_i(t)}(dx^i)^2, \qquad (5)$$

where we imply summing up over space index $i=1,...,n$, and on the metric are imposed the conditions

$$\sum_{i=1}^{n}\beta_i = 0,$$
$$\frac{1}{n(n-1)}\sum_{i=1}^{n}\dot\beta_i^2 = \dot\alpha^2; \qquad (6)$$

dot above the letter denotes derivative by time. We'll consider first conditions (6) as valid whole the time, second condition in (6) usually is breaking with presence of matter or vacuum energy. Disturbed metric coefficient we'll write down in the form

$$g_{\mu\nu}(t) + h_{\mu\nu}(x) = e^{2\alpha(t)+2\beta_i(t)+2\xi_i(x)} \approx e^{2\alpha(t)+2\beta_i(t)}(1+2\xi_i(x)),$$

where is no summation; so we get

$$h_{ii}(x) = 2e^{2\alpha(t)+2\beta_i(t)}\xi_i(x);$$

Let's think that perturbations of metric don't lead us off the original anisotropic Kazner model, that is first equation in (6) gives

$$\sum_{i=1}^{n}\xi_i = 0. \qquad (7)$$

From here

$$e^{-2\beta_i(t)}h_{ii}(x) = h_i^i = h = 0, \qquad (8)$$

where's implying summation over index $i=1,...,n$. Thus we have

$$h = \bar{h} = 0,$$
$$\bar{h}_{\mu\nu} = h_{\mu\nu}. \qquad (9)$$

Let's impose additional conditions

$$h_{0\mu} = 0;$$

in reality we're using here standard TT-gauge, which uses ordinary in linearized theory in case of three-dimensional space-hypersurface [6].
Eq. (4) we could rewrite down now, taking into account Eq. (3), (9) in form

$$h_{\mu\nu|\alpha}{}^{\alpha} + 2R_{\alpha\mu\beta\nu}h^{\alpha\beta} - \frac{6\Lambda}{n-1}h_{\mu\nu} = 0, \tag{10}$$

it'll be the main for the gravitational perturbations in the model.
The gauge Lorentz condition (2) with using Eq. (8) gives us

$$e^{-2\beta_i}\dot{\beta}_i h_{ii} = 0, \tag{11}$$

where is using summation over space indexes.
Einstein's Eq. (3) for the case of undisturbed metric with cosmological constant have the form

$$\dot{\alpha}^2 = \frac{1}{n(n-1)}\left(2\Lambda + \sum_i \dot{\beta}_i^2\right) \tag{12}$$

$$\ddot{\beta}_j + n\dot{\alpha}\dot{\beta}_j = 0$$

If density of energy of anisotropy $\rho_{anis}$ is more than density of energy of matter, but here that is vacuum energy $\rho_{vac}$,

$$\rho_{vac} \ll \frac{c^2}{16\pi G}\sum_i \dot{\beta}_i^2 = \rho_{anis}, \tag{13}$$

first Einstein's Eq. (12) is coinciding with the second Kazner condition (6).
Decisions of system Eq. (12) for the case of positive cosmological term are [15]

$$\alpha(t) = \frac{1}{n}\ln\left(\frac{B}{\sqrt{2\Lambda}}sh\lambda t\right), \quad \beta_k = \frac{B_k}{B}\sqrt{\frac{n-1}{n}}\ln\left(\frac{B}{\sqrt{2\Lambda}}th(\lambda t/2)\right),$$

$$\lambda = \sqrt{\frac{2\Lambda n}{n-1}}, \quad \Lambda > 0;$$
(14)

where

$$\dot{\beta}_k = B_k e^{-n\alpha}, \quad B^2 \equiv \sum_{k=1}^{n} B_k^2, \quad B_k = const. \tag{15}$$

Expressions (14) will be considered as background metric for the main equation of propagating of gravitational waves (4).

For the early stages of evolution, with small times when $\lambda t \ll 1$, decision's representing by ordinary Kazner metric with conditions (6); when we investigate long times $\lambda t \gg 1$, decision's transforming into exponential extension of de-Sitter, when cosmological constant $\Lambda$ is prevailing.

Non-zero's Christoffel symbols and components of Riemann curvature tensor for the diagonal metric (5) have the form [15]

$$\Gamma^i_{i0} = \dot{\alpha} + \dot{\beta}_i, \quad \Gamma^0_{ii} = (\dot{\alpha} + \dot{\beta}_i)e^{2(\alpha+\beta_i)}, \tag{16}$$

$$R^{\hat{i}}_{\hat{j}\hat{i}\hat{j}} = (\dot{\alpha} + \dot{\beta}_i)(\dot{\alpha} + \dot{\beta}_j), \quad i \neq j;$$

little hat over indexes denotes orthonormal system of coordinate with basic 1-forms [15]:

$$\omega^{\hat{0}} = dt, \quad \omega^{\hat{i}} = dx^i,$$

$$ds^2 = -(\omega^{\hat{0}})^2 + (\omega^{\hat{1}})^2 + \ldots + (\omega^{\hat{n}})^2.$$

Calculating covariant derivatives, and taking into account Eq. (15), (16), (8), (11) we get for the terms of Eq. (10):

$$h_{kj|\alpha}{}^{|\alpha} = -\ddot{h}_{kj} + e^{-2(\alpha+\beta_i)}h_{kj,ii} + \dot{h}_{kj}((4-n)\dot{\alpha} + 2(B_i + B_j)e^{-n\alpha}) + h_{kj}(2\ddot{\alpha} + 2(n-2)\dot{\alpha}^2$$

$$-4(B_k + B_j)\dot{\alpha}e^{-n\alpha} - (B_k + B_j)^2 e^{-2n\alpha}) \approx -\ddot{h}_{kj} + (4-n)\sqrt{\frac{2\Lambda}{n(n-1)}}\dot{h}_{kj} + \frac{4(n-2)}{n(n-1)}\Lambda h_{kj}; \tag{17}$$

$$2R_{\alpha j\beta j}h^{\alpha\beta} = -2h_{jj}(\dot{\alpha} + \dot{\beta}_j)^2 = -2h_{jj}(\dot{\alpha} + B_j e^{-n\alpha})^2 \approx -\frac{4}{n(n-1)}\Lambda h_{jj}, \tag{18}$$

$$R_{\alpha k\beta j}h^{\alpha\beta} = R_{\alpha 0\beta\mu}h^{\alpha\beta} = 0, \quad k \neq j;$$

here're remained only main terms of asymptotic for the expressions when $\lambda t \gg 1$. Substituting expressions (17), (18) into Eq. (10), we get

$$\ddot{h}_{jj} + (n-4)\sqrt{\frac{2\Lambda}{n(n-1)}}\dot{h}_{jj} + \frac{2(n+6)}{n(n-1)}\Lambda h_{jj} = 0,$$

$$\ddot{h}_{kj} + (n-4)\sqrt{\frac{2\Lambda}{n(n-1)}}\dot{h}_{kj} + \frac{2(n+4)}{n(n-1)}\Lambda h_{kj} = 0, \quad k \neq j. \tag{19}$$

It could be written down as ordinary equations for harmonic oscillator:

$$\ddot{h}_{jj} + 2\delta \dot{h}_{jj} + \Omega^2 h_{kj} = 0,$$

(20)

$$\ddot{h}_{kj} + 2\delta \dot{h}_{kj} + \Omega_1^2 h_{kj} = 0, \quad k \neq j,$$

where

$$\delta = (n-4)\sqrt{\frac{\Lambda}{2n(n-1)}}, \quad \Omega^2 = \frac{2(n+6)}{n(n-1)}\Lambda, \quad \Omega_1^2 = \frac{2(n+4)}{n(n-1)}\Lambda.$$

Decisions of Eq. (20) could be written down right away, for example for diagonal components we take $h_{jj} = H_{jj}e^{\eta t}$ and get

$$\eta = -\delta \pm \sqrt{\delta^2 - \Omega^2} = \frac{\lambda}{2n}(4 - n \pm \sqrt{(n-6)^2 - 44}).$$

If $\delta > \Omega$, that corresponds $n > 6 + \sqrt{44} \approx 12$, than decision has the form of aperiodic fading

$$h_{jj} = e^{-\delta t}(H_{1jj}e^{\sqrt{\delta^2 - \Omega^2}t} + H_{2jj}e^{-\sqrt{\delta^2 - \Omega^2}t}), \qquad (22)$$

If $\delta < \Omega$, that corresponds $n = 3,...,12$, than

$$\eta = -\delta \pm i\sqrt{\Omega^2 - \delta^2} = \frac{\lambda}{2n}(4 - n \pm i\sqrt{44 - (n-6)^2}),$$

decision could be represented as oscillation

$$h_{jj} = H_{jj}e^{-\delta t}Sin(\omega t + \varphi_j), \quad \omega = \frac{\lambda}{2n}\sqrt{44 - (n-6)^2}, \quad \delta = \frac{\lambda}{2n}(n-4), \qquad (23)$$

where $H_{jj}, \varphi_j$ - initial amplitudes and phases of disturbances respectively. Further we'll consider that $\varphi_j = 0$.

Thus when number of space dimensions equals $n = 3$, perturbation increases during the time, $\delta < 0$, that is background decision (14) is unstable. When $n = 4$ perturbation has purely fadeless oscillating character, $\delta = 0$. When $5 \leq n \leq 12$ perturbation has the form of fading oscillations with $\delta > 0$. When $n \geq 13$ perturbation's aperiodically fading in time, $\delta > \Omega$. Multidimensional Kazner decision (14) is asymptotically stable only for the number of space dimensions $n \geq 4$.

Decision for nondiagonal metric components could be obtained by the same manner. They'll have the same form as (22), (23), but we should use in (23) instead of $\omega$ new expression for the frequency $\omega_1$,

$$h_{jk} = H_{jk}e^{-\delta t}Sin(\omega_1 t + \theta_{jk}), \quad \omega_1 = \frac{\lambda}{2}\sqrt{\frac{12-n}{n}}, \quad \delta = \frac{\lambda}{2n}(n-4),, \qquad (24)$$

where $H_{jk}, \theta_{jk}$ - initial amplitudes and phases. Here we'll consider that $\theta_{jk} = 0$.

Critical dimension for gravitational oscillations of metric remains the same, $n = 12$. Usual oscillations of both types of diagonal and nondiagonal metric components are available only in the space maximum with $n = 11$.

Relict long-period oscillations of metric, which are interested in the model, were borning as fluctuations in early moments of time, and would affect to the dynamic of expanding Universe in the late moments of time. Let's consider question about energy-momentum tensor for relict gravitational waves. We'll think that metric perturbation arose at the early moments of creating the metric, when was $\lambda t \approx 0$. Using (21), we have [6]

$$T^{GW}_{\mu\nu} = \frac{1}{32\pi} <h_{\alpha\beta|\mu} h^{\alpha\beta}{}_{|\nu}>. \tag{25}$$

Here angular brackets denote averaging along some wave length, but because we've neglected dependence of oscillation from space coordinates in Eq. (17), for the density of energy of gravitational oscillations in case of $3 \leq n \leq 12$

$$T^{GW}_{00} = \frac{1}{32\pi} h_{\alpha\beta|0} h^{\alpha\beta}{}_{|0} = \frac{1}{32\pi} h_{ik|0} h^{ik}{}_{|0} = \frac{1}{32\pi} h_{ii|0} h^{ii}{}_{|0} + \frac{1}{32\pi} h_{jk|0} h^{jk}{}_{|0}, \tag{26}$$

where $j \neq k$. We have two contributions into the density of energy

$$T^{diag}_{00} = \frac{1}{32\pi} h_{ii|0} h^{ii}{}_{|0} = \frac{1}{32\pi} e^{-\lambda t} \sum_{i=1}^{n} \left(\frac{H_{ii}}{d_i^2}\right)^2 (\omega Cos\omega t - (\delta + 2\dot{\alpha} + 2\dot{\beta}_i) Sin\omega t)^2,$$

$$T^{ndiag}_{00} = \frac{1}{32\pi} h_{jk|0} h^{jk}{}_{|0} = \frac{1}{32\pi} e^{-\lambda t} \sum_{j \neq k}^{n} \left(\frac{H_{jk}}{d_j d_k}\right)^2 (\omega_1 Cos\omega_1 t - (\delta + 2\dot{\alpha} + \dot{\beta}_i + \dot{\beta}_k) Sin\omega_1 t)^2.$$

Here were used asymptotics for $\lambda t \gg 1$

$$e^{-(\alpha + \beta_i)} \approx \frac{e^{-\frac{\lambda t}{n}}}{d_i}, \quad d_i = 2^{-\frac{1}{n}} \left(\frac{B}{\sqrt{2\Lambda}}\right)^{1/n + C_i}, \quad C_i = \frac{B_i}{B}\sqrt{\frac{n-1}{n}}.$$

For the $T^{GW}_{00}$ we get for contributions of diagonal and nondiagonal terms into the density of energy

$$T^{diag}_{00} = \frac{1}{32\pi} e^{-\lambda t} \sum_{i=1}^{n} \left(\frac{H_{ii}}{d_i^2}\right)^2 (\omega Cos\omega t - (\delta + \dot{\alpha}) Sin\omega t)^2 =$$

$$= \frac{1}{32\pi} e^{-\lambda t} \sum_{i=1}^{n} \left(\frac{H_{ii}}{d_i^2}\right)^2 (\frac{\lambda}{2n}\sqrt{44 - (n-6)^2} Cos\omega t - \frac{\lambda}{2} Sin\omega t)^2 =$$

$$= \frac{3n+2}{16\pi n(n-1)} \Lambda e^{-\lambda t} \sum_{i=1}^{n} \left(\frac{H_{ii}}{d_i^2}\right)^2 Sin^2(\omega t - \varphi),$$

where $tg\varphi = \dfrac{\sqrt{44-(n-6)^2}}{n}$;

$$T_{00}^{ndiag} = \dfrac{1}{32\pi}e^{-\lambda t}\sum_{j\neq k}\left(\dfrac{H_{jk}}{d_j d_k}\right)^2 (\omega_1 Cos\omega_1 t - (\delta+\alpha)Sin\omega_1 t)^2 =$$

$$= \dfrac{1}{32\pi}e^{-\lambda t}\sum_{j\neq k}\left(\dfrac{H_{jk}}{d_j d_k}\right)^2 (\dfrac{\lambda}{2}\sqrt{\dfrac{12-n}{n}}Cos\omega_1 t - \dfrac{\lambda}{2}Sin\omega_1 t)^2 =$$

$$= \dfrac{3}{16\pi n(n-1)}\Lambda e^{-\lambda t}\sum_{j\neq k}\left(\dfrac{H_{jk}}{d_j d_k}\right)^2 Sin^2(\omega_1 t - \varphi_1),$$

where $tg\varphi_1 = \sqrt{\dfrac{12-n}{n}}$.

Finally we get

$$T_{00}^{GW} = T_{00}^{diag} + T_{00}^{ndiag} = \dfrac{\Lambda e^{-\lambda t}}{16\pi n(n-1)}((3n+2)\sum_{i=1}^{n}\left(\dfrac{H_{ii}}{d_i^2}\right)^2 Sin^2(\omega t-\varphi) +$$

(27)

$$+ 3\sum_{j\neq k}\left(\dfrac{H_{jk}}{d_j d_k}\right)^2 Sin^2(\omega_1 t - \varphi_1)).$$

We've got slow oscillations of expanding metric which possess density of energy. That density needs to be taking into account in the right side of Einstein's Eq. (3) for background metric and to consider it like addition to the vacuum energy. Naturally, such oscillations change depending on time vacuum of the model, so when we have to consider quantum fields, for instance scalar field [16], there'll be presented additional contribution into cosmological arising of particles. With such accuracy energy-momentum tensor of gravitational oscillations coming out just like energy-momentum tensor of matter. It plays the same role in creating of background curvature and in the same form should be appeared in law of conservation just as ordinary energy-momentum tensor of matter.

Value of $T_{00}^{GW}$ have to satisfy the restriction for the energy-momentum tensor of matter:

$$T_{00}^{GW} \ll \dfrac{1}{16\pi}\sum_i \dot{\beta}_i^2 = \dfrac{B^2}{16\pi}e^{-2n\alpha(t)},$$

or

$$\dfrac{1}{8n(n-1)}((3n+2)\sum_{i=1}^{n}\left(\dfrac{H_{ii}}{d_i^2}\right)^2 Sin^2(\omega t-\varphi) +$$

$$+ 3\sum_{j\neq k}\left(\dfrac{H_{jk}}{d_j d_k}\right)^2 Sin^2(\omega_1 t - \varphi_1)) \ll e^{-\lambda t},$$

for long times this condition is not valid, so there exists some restriction on time for using of background decision (14). At some moment of time the energy of gravitational oscillations will be comparable with the energy of anisotropy, and this demonstrates the necessity of using it when we want to calculate de-Sitter stage of expanding. Part of those energy could be sufficiently big for long times and play important role along with $\Lambda$.

It's interesting to compare energy of proper background oscillations with the density of energy of «vacuum», conditioned by particles arising during the expansion. In case of mass scalar field, that value is given by expression [16]

$$\rho \approx \frac{1}{8(2\pi)^n} \frac{\sqrt{2\Lambda}}{B} \frac{\exp(-(1+4/n)\lambda t)D_k}{(m^2 - \lambda^2/4)(m^2 - \lambda^2/4 + \lambda^2/n^2)}, \qquad (28)$$

where $D_k$ - constant, specified by offcut integral over momentum. Compared that with Eq. (27) for $T_{00}^{GW}$, could be seen that $T_{00}^{GW}$ is decreasing slowly than $\rho_{vac}$ with time, that is in Einstein's equation

$$R_{\alpha\beta} - \frac{1}{2}g_{\alpha\beta}R = -\Lambda g_{\alpha\beta} + 8\pi(T_{\alpha\beta}^{GW} + T_{\alpha\beta}^{vac}) \qquad (29)$$

in the right side in second approximation we could use only cosmological term with $\Lambda$ and $T_{00}^{GW}$, surely it concerns only latest Universe. Also in the right side of Eq. (3) we could put down energy-momentum tensor, that is in charge of arising of particles caused by proper oscillations of background, but it's contribution should be more less in compare with $T_{\alpha\beta}^{GW}$ and $T_{\alpha\beta}^{vac}$. That contribution could be evaluated by famous method used in [16], and as it turned out, it contains additional factor $e^{-\frac{\lambda}{2}t}$ compared with expression (28), and we could neglect it in our approximation.

Thus we've got the most evident contributions into energy-momentum tensor of relict gravitational waves in the late stage of expansion. The next step is getting a little bit more accurate decision of Einstein's Eq. (3) with contribution of gravitational waves (27), especially on the late stages of evolution of Universe, when density of energy of oscillation is becoming comparable with the $\Lambda$. In that way energy of metric oscillations could give contributions into so called «dark energy».

As far as we know from the theory of early Universe, value of fluctuations in the beginning could be expected as order as $10^{-5}$. That fluctuations could be developed only in the world with three space dimensions. When number of space dimensions $4 \leq n \leq 12$, fluctuations have oscillating character, and are fading aperiodically when $n > 12$.